\documentclass[12pt,a4paper]{article}

\usepackage{lineno}  % for line numbering during review
\usepackage{graphicx}  % to include figures (can also use other packages)
\usepackage{xspace} % To avoid problems with missing or double spaces after
\usepackage{color}
\usepackage{colortbl}
\usepackage{amsmath} % Adds a large collection of math symbols
\usepackage{ifthen} % for conditional statements
\graphicspath{{./figs/}} % Make Latex search fig subdir for figures

\newboolean{pdflatex}
\setboolean{pdflatex}{true} % use this if using non-eps figures
\newboolean{articletitles}
\setboolean{articletitles}{true} % False removes titles in references

\newboolean{uprightparticles}
\setboolean{uprightparticles}{false} %Set to true to get roman particle symbols

\usepackage{amssymb}
\usepackage{amsfonts}
\usepackage{upgreek} % Adds in support for greek letters in roman typeset

\usepackage{hyperref}    % Hyperlinks in references
\usepackage[all]{hypcap} % Internal hyperlinks to floats.

\def\lhcb {\mbox{LHCb}\xspace}

\def\MagUp {\mbox{\em Mag\kern -0.05em Up}\xspace}

\ifthenelse{\boolean{uprightparticles}}%
{

 \def\PDelta      {\ensuremath{\Delta}\xspace}                 
 \def\PXi      {\ensuremath{\Xi}\xspace}                 
 \def\PLambda      {\ensuremath{\Lambda}\xspace}                 
 \def\PSigma      {\ensuremath{\Sigma}\xspace}                 
 \def\POmega      {\ensuremath{\Omega}\xspace}                 
 \def\PUpsilon      {\ensuremath{\Upsilon}\xspace}                 
 
 \def\PB      {\ensuremath{\mathrm{B}}\xspace}                 
                  
 \def\PD      {\ensuremath{\mathrm{D}}\xspace}

 \def\PK      {\ensuremath{\mathrm{K}}\xspace}

 \def\Pb      {\ensuremath{\mathrm{b}}\xspace}                 
 \def\Pc      {\ensuremath{\mathrm{c}}\xspace}

 \def\Pi      {\ensuremath{\mathrm{i}}\xspace}

}
{

 \mathchardef\PDelta="7101
 \mathchardef\PXi="7104
 \mathchardef\PLambda="7103
 \mathchardef\PSigma="7106
 \mathchardef\POmega="710A
 \mathchardef\PUpsilon="7107
                  
 \def\PB      {\ensuremath{B}\xspace}                 
                  
 \def\PD      {\ensuremath{D}\xspace}

 \def\PK      {\ensuremath{K}\xspace}

 \def\Pb      {\ensuremath{b}\xspace}                 
 \def\Pc      {\ensuremath{c}\xspace}

 \def\Pi      {\ensuremath{i}\xspace}

}

\makeatletter
\ifcase \@ptsize \relax% 10pt
  \newcommand{\miniscule}{\@setfontsize\miniscule{4}{5}}% \tiny: 5/6
\or% 11pt
  \newcommand{\miniscule}{\@setfontsize\miniscule{5}{6}}% \tiny: 6/7
\or% 12pt
  \newcommand{\miniscule}{\@setfontsize\miniscule{5}{6}}% \tiny: 6/7
\fi
\makeatother

\DeclareRobustCommand{\optbar}[1]{\shortstack{{\miniscule (\rule[.5ex]{1.25em}{.18mm})}
  \\ [-.7ex] $#1$}}

\def\cquark    {{\ensuremath{\Pc}}\xspace}

\def\bquark    {{\ensuremath{\Pb}}\xspace}

  \def\Kbar    {{\kern 0.2em\overline{\kern -0.2em \PK}{}}\xspace}

\def\KorKbar    {\kern 0.18em\optbar{\kern -0.18em K}{}\xspace}

  \def\Dbar    {{\kern 0.2em\overline{\kern -0.2em \PD}{}}\xspace}

\def\DorDbar    {\kern 0.18em\optbar{\kern -0.18em D}{}\xspace}

\def\Bbar    {{\ensuremath{\kern 0.18em\overline{\kern -0.18em \PB}{}}}\xspace}

\def\BorBbar    {\kern 0.18em\optbar{\kern -0.18em B}{}\xspace}

  \def\Y#1S{\ensuremath{\PUpsilon{(#1S)}}\xspace}% no space before {...}!

\def\Lbar        {{\ensuremath{\kern 0.1em\overline{\kern -0.1em\PLambda}}}\xspace}
\def\LorLbar    {\kern 0.18em\optbar{\kern -0.18em \PLambda}{}\xspace}

\def\eps   {{\ensuremath{\varepsilon}}\xspace}

\def\AT#1     {\ensuremath{A_{\mathrm{T}}^{#1}}\xspace}           % 2

\def\C#1      {\ensuremath{\mathcal{C}_{#1}}\xspace}                       % 9
\def\Cp#1     {\ensuremath{\mathcal{C}_{#1}^{'}}\xspace}                    % 7
\def\Ceff#1   {\ensuremath{\mathcal{C}_{#1}^{\mathrm{(eff)}}}\xspace}        % 9  
\def\Cpeff#1  {\ensuremath{\mathcal{C}_{#1}^{'\mathrm{(eff)}}}\xspace}       % 7
\def\Ope#1    {\ensuremath{\mathcal{O}_{#1}}\xspace}                       % 2
\def\Opep#1   {\ensuremath{\mathcal{O}_{#1}^{'}}\xspace}                    % 7

\newcommand{\tev}{\ifthenelse{\boolean{inbibliography}}{\ensuremath{~T\kern -0.05em eV}\xspace}{\ensuremath{\mathrm{\,Te\kern -0.1em V}}}\xspace}
\newcommand{\gev}{\ensuremath{\mathrm{\,Ge\kern -0.1em V}}\xspace}
\newcommand{\mev}{\ensuremath{\mathrm{\,Me\kern -0.1em V}}\xspace}
\newcommand{\kev}{\ensuremath{\mathrm{\,ke\kern -0.1em V}}\xspace}
\newcommand{\ev}{\ensuremath{\mathrm{\,e\kern -0.1em V}}\xspace}
\newcommand{\gevc}{\ensuremath{{\mathrm{\,Ge\kern -0.1em V\!/}c}}\xspace}
\newcommand{\mevc}{\ensuremath{{\mathrm{\,Me\kern -0.1em V\!/}c}}\xspace}
\newcommand{\gevcc}{\ensuremath{{\mathrm{\,Ge\kern -0.1em V\!/}c^2}}\xspace}
\newcommand{\gevgevcccc}{\ensuremath{{\mathrm{\,Ge\kern -0.1em V^2\!/}c^4}}\xspace}
\newcommand{\mevcc}{\ensuremath{{\mathrm{\,Me\kern -0.1em V\!/}c^2}}\xspace}

\def\mum  {\ensuremath{{\,\upmu\rm m}}\xspace}

\def\gsim{{~\raise.15em\hbox{$>$}\kern-.85em
          \lower.35em\hbox{$\sim$}~}\xspace}
\def\lsim{{~\raise.15em\hbox{$<$}\kern-.85em
          \lower.35em\hbox{$\sim$}~}\xspace}

\def\ptot       {\mbox{$p$}\xspace}
\def\pt         {\mbox{$p_{\rm T}$}\xspace}

\def\evtgen     {\mbox{\textsc{EvtGen}}\xspace}

\def\geant      {\mbox{\textsc{Geant4}}\xspace}

\def\photos     {\mbox{\textsc{Photos}}\xspace}

\def\pythia     {\mbox{\textsc{Pythia}}\xspace}

\def\tell1  {TELL1\xspace}
\def\ukl1   {UKL1\xspace}

\usepackage{cite} % Allows for ranges in citations
\usepackage{mciteplus}
\usepackage{slashbox}

\begin{document}

%%%%%%%%%%%%%%%%%%%%%%%%%
%%%%% Title     %%%%%%%%%
%%%%%%%%%%%%%%%%%%%%%%%%%
\renewcommand{\thefootnote}{\fnsymbol{footnote}}
\setcounter{footnote}{1}

% %%%%%%% CHOOSE --------
% \input{title-LHCb-ANA}
% \input{title-LHCb-CONF}
% $Id: title-LHCb-PAPER.tex 56951 2014-06-30 13:45:01Z roldeman $
% ===============================================================================
% Purpose: LHCb-PAPER journal paper title page template
% Author: 
% Created on: 2010-09-25
% ===============================================================================

%%%%%%%%%%%%%%%%%%%%%%%%%
%%%%%  TITLE PAGE  %%%%%%
%%%%%%%%%%%%%%%%%%%%%%%%%
\begin{titlepage}
\pagenumbering{roman}

% Header ---------------------------------------------------
\vspace*{-1.5cm}
\centerline{\large EUROPEAN ORGANIZATION FOR NUCLEAR RESEARCH (CERN)}
\vspace*{1.2cm}
\hspace*{-0.5cm}
\begin{tabular*}{\linewidth}{lc@{\extracolsep{\fill}}r}
\ifthenelse{\boolean{pdflatex}}% Logo format choice
{\vspace*{-3.7cm}\mbox{\!\!\!\includegraphics[width=.14\textwidth]{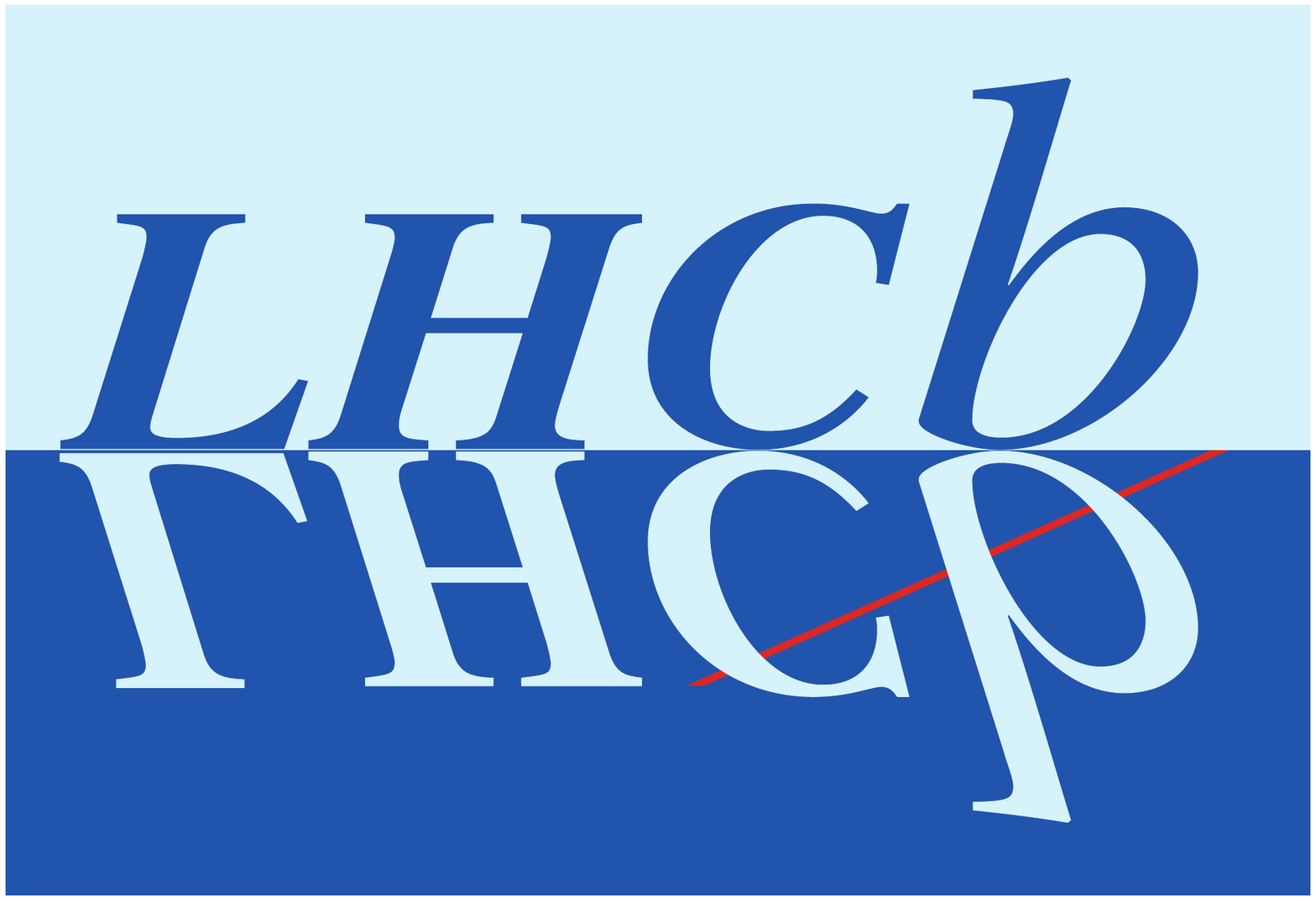}} & &}%
{\vspace*{-1.2cm}\mbox{\!\!\!\includegraphics[width=.12\textwidth]{lhcb-logo.eps}} & &}%
\\
 & & CERN-PH-EP-2014-285 \\  % ID 
 & & LHCb-PAPER-2014-057 \\  % ID 
 & & December 6, 2014 \\ % Date - Can also hardwire e.g.: 23 March 2010
%% & & Version 4.0 \\
% not in paper \hline
\end{tabular*}

\vspace*{2.0cm}

% Title --------------------------------------------------
{\bf\boldmath\huge
\begin{center}
Measurement of the inelastic $pp$\/ cross-section 
at a centre-of-mass energy of $\sqrt{s}=7$\,TeV
\end{center}
}

\vspace*{2.0cm}

% Authors -------------------------------------------------
\begin{center}
%In the footnote, replace 'paper' by 'letter' in case of submission to PRL or PLB 
The LHCb collaboration\footnote{Authors are listed at the end of this paper.}
\end{center}

\vspace{\fill}

% Abstract -----------------------------------------------
\begin{abstract}
  \noindent
The cross-section for inelastic proton-proton collisions, with at 
least one prompt long-lived charged particle of transverse momentum 
$\pt>0.2\,\gevc$\/ in the pseudorapidity range $2.0<\eta<4.5$, 
is measured by the LHCb experiment at a centre-of-mass energy of 
$\sqrt{s}=7$\,TeV. The cross-section in this kinematic range is 
determined to be $\sigma_{\rm inel}^{\rm acc} = 55.0\,\pm\,2.4$\,mb
with an experimental uncertainty that is dominated by systematic contributions.
Extrapolation to the full phase space, using \pythia{\,}6, yields 
$\sigma_{\rm inel} = 66.9 \,\pm\,2.9\,\pm\,4.4$\,mb, where the first
uncertainty is experimental and the second is due to the extrapolation.
\end{abstract}

\vspace*{2.0cm}

\begin{center}
  Submitted to JHEP
\end{center}

\vspace{\fill}

{\footnotesize 
\centerline{\copyright~CERN on behalf of the \lhcb collaboration, license \href{http://creativecommons.org/licenses/by/4.0/}{CC-BY-4.0}.}}
\vspace*{2mm}

\end{titlepage}

%%%%%%%%%%%%%%%%%%%%%%%%%%%%%%%%
%%%%%  EOD OF TITLE PAGE  %%%%%%
%%%%%%%%%%%%%%%%%%%%%%%%%%%%%%%%

%  empty page follows the title page ----
\newpage
\setcounter{page}{2}
\mbox{~}
%\newpage
%
%% Author List ----------------------------
%%  You need to get a new author list!
%\input{LHCb_authorlist.tex}
%\input{LHCb_HD_authorlist_2014-09-26.tex}
%
%The author list for journal publications is provided by the Membership Committee shortly after 'approval to go to paper' has been given.
%%It will be made available on the page
%%\verb!http://www.physik.uzh.ch/~strauman/forMemCo/LHCb-PAPER-XXXX-XXX/! .
%It will be sent to you by email shortly after a paper number has beens assigned.
%The author list should be included already at first circulation, 
%to allow new members of the collaboration to verify whether they have been included correctly.
%Occasionally a misspelled name is corrected or associated institutions become full members.
%In that case, a new author list will be sent to you.
%In case line numbering doesn't work well after including the authorlist, try moving the \verb!\bigskip! after the last author to a separate line.
%
%
%The authorship for Conference Reports should be ``The LHCb
%  collaboration'', with a footnote giving the name(s) of the contact
%  author(s), but without the full list of collaboration names.

\cleardoublepage

% %%%%%%%%%%%%% ---------

\renewcommand{\thefootnote}{\arabic{footnote}}
\setcounter{footnote}{0}

%%%%%%%%%%%%%%%%%%%%%%%%%%%%%%%%
%%%%%  Table of Content   %%%%%%
%%%%%%%%%%%%%%%%%%%%%%%%%%%%%%%%
%%%% Uncomment next 2 lines if desired
%\tableofcontents
%\cleardoublepage

%%%%%%%%%%%%%%%%%%%%%%%%%
%%%%% Main text %%%%%%%%%
%%%%%%%%%%%%%%%%%%%%%%%%%

\pagestyle{plain} % restore page numbers for the main text
\setcounter{page}{1}
\pagenumbering{arabic}

% %%%%%%% CHOOSE --------
%% ----------------------------------
%% Line numbering on the left margin 
%% ----------------------------------
%% Uncomment during review phase. 
%% Comment it out before a final submission.
%%%%%%%%%%%%%%\linenumbers
%% --------------------------------
% %%%%%%%%%%%%% ---------

%% here comes the actual paper  

\section{Introduction}
%----------------------
\label{sec:Introduction}
The inelastic cross-section is a fundamental observable in 
high-energy hadronic interactions. It is also important in 
astroparticle physics for models of extensive air showers 
induced by cosmic rays in the atmosphere \cite{Knapp:1997bb}. Currently, 
it is not possible to calculate its value from first principles because 
quantum chromodynamics cannot yet be solved for soft processes.
Phenomenological models assume a rise of the inelastic cross-section 
with energy according to a power law \cite{Sjostrand:2006za,Sjostrand:2008za}, 
while not exceeding the Froissart-Martin bound \cite{Froissart:1961,Martin:1966}, 
which is asymptotically proportional to $\ln^2s$. Although originally 
the Froissart-Martin bound was derived for the total cross-section, 
later developments show that it is also valid for the inelastic 
cross-section \cite{Martin:2009}. 
 
Measurements of the inelastic proton-proton ($pp$) cross-section at 
$\sqrt{s}=7$\,TeV have been reported by the ALICE \cite{ALICEcs:2013}, 
ATLAS \cite{ATLAScs:2011,ATLASxs:2014}, CMS \cite{CMScs:2011} and 
TOTEM  \cite{TOTEMcs:2011,Antchev:2013iaa} collaborations, using 
experimental information from the central (ALICE, ATLAS, CMS) and 
the extremely forward (ATLAS, TOTEM) regions. LHCb allows those results 
to be complemented by a measurement in the mid- to forward rapidity
range $2.0<\eta<4.5$.

\section{Detector description and data set}
%------------------------------------------
\label{sec:Detector}
The \lhcb detector~\cite{Alves:2008zz} is a single-arm forward
spectrometer covering the \mbox{pseudorapidity} range $2<\eta <5$,
designed for the study of particles containing \bquark or \cquark
quarks. The detector includes a high-precision tracking system
consisting of a silicon-strip vertex detector surrounding the $pp$\/
interaction region~\cite{LHCbVELOGroup:2014uea}, a large-area 
silicon-strip detector located upstream of a dipole magnet with 
a bending power of about $4{\rm\,Tm}$, the polarity of which 
can be inverted, and three stations of 
silicon-strip detectors and straw drift tubes~\cite{LHCb-DP-2013-003} 
placed downstream of the magnet. The tracking system provides a 
measurement of momentum, \ptot, with a relative uncertainty that 
varies from 0.4\% at low momentum to 0.6\% at 100\gevc. The 
minimum distance of a track to a primary vertex, the impact parameter, 
is measured with a resolution of $(15+29/\pt)\mum$, where \pt is 
the component of the momentum transverse to the beam, in \gevc. Different 
types of charged hadrons are distinguished using information from 
two ring-imaging Cherenkov detectors. Photon, electron and hadron 
candidates are identified by a calorimeter system consisting of 
scintillating-pad and preshower detectors, an electromagnetic 
calorimeter and a hadronic calorimeter. Muons are identified by a 
system composed of alternating layers of iron and multiwire 
proportional chambers. The trigger~\cite{LHCb-DP-2012-004} consists 
of a hardware stage, based on information from the calorimeter and 
muon systems, followed by a software stage, which applies a full 
event reconstruction. 

In the simulation, $pp$\/ collisions are generated using
\pythia 6 \cite{Sjostrand:2006za} with a specific \lhcb
configuration \cite{LHCb-PROC-2010-056} using the CTEQ\,6 leading-order 
parton density functions. Decays of hadronic particles
are described by \evtgen \cite{Lange:2001uf}, in which final-state
radiation is generated using \photos \cite{Golonka:2005pn}. The
interaction of the generated particles with the detector, and its
response, are implemented using the \geant\/ toolkit 
\cite{Allison:2006ve, Agostinelli:2002hh} as described in
Ref.~\cite{LHCb-PROC-2011-006}.

The data used in this analysis are a subset of the data recorded 
during low-luminosity running in early 2010 with a minimum bias trigger 
where the hardware stage triggered every beam-beam crossing and the 
event was accepted at the software stage if at least one reconstructed track 
segment was found in the vertex detector. Using a sample of no-bias triggered events, 
it has been checked that for the events selected in this analysis, the 
trigger efficiency exceeds 99.99\%. From the rate of empty events 
the average number of interactions per bunch crossing, $\mu$, with at 
least one track in the detector, was estimated to be 0.1. This corresponds 
to $P=\mu/(1-\exp(-\mu)) \approx 1.05$\/ visible interactions per 
triggered event. The measurement is based on integrated luminosities of 
0.62 (1.25)\,nb$^{-1}$\/ recorded with the magnetic field 
polarity in the upward (downward) direction. The integrated luminosity 
has been determined with an overall precision of 3.5\% 
\cite{LHCb-PAPER-2011-015}.

\section{Data analysis}
%----------------------
\label{sec:analysis}
This analysis measures the inelastic $pp$\/ cross-section for the
production of at least one prompt long-lived charged particle 
with $\pt>0.2$\,\gevc and pseudorapidity in the range $2.0<\eta<4.5$. 
A prompt particle is defined as one whose impact parameter relative 
to the point of the primary interaction is smaller than 200\,\mum.
   
The LHCb coordinate system is a right-handed cartesian system with the $z$\/ 
axis along the average beam direction from the vertex detector towards 
the muon system, the $y$\/ axis pointing upward and $x$\/ 
towards the outside of the LHC. Reconstructed tracks are required to 
have a track segment in the vertex detector and in the tracking system 
downstream of the magnet. Selection criteria (cuts) are applied on the 
track fit $\chi^2/{\rm NDF}$, with ${\rm NDF}$\/ the number of degrees 
of freedom of the fit, and on the distance of closest approach, 
${\rm DCA}$, to the longitudinal axis of the luminous region. This 
axis is determined by the mean values of Gaussian functions fitted 
in bins of $z$\/ to the $x$\/ and $y$\/ distributions of reconstructed 
primary vertices. To suppress background from beam-gas interactions, 
the $z$\/ coordinate of the midpoint between the points of closest 
approach on the reconstructed particle trajectory and on the 
longitudinal axis of the luminous region is required to satisfy 
$|z-z_c|<130$\,mm. Here $z_c$\/ is the longitudinal centre of the luminous 
region, determined by the mean value of a Gaussian function fitted to 
the $z$\/ distribution of the reconstructed primary vertices. The width
of the distribution is found to be $\sigma_z = 38.2$\,mm. The determination
of the central axis of the luminous region and its longitudinal centre
is done separately for each magnet polarity. The analysis is restricted 
to tracks in a fiducial region away from areas where the magnetic field or
detector geometry cause sharp variations in the track finding efficiency. 

The cross-section, $\sigma_{\rm inel}^{\rm acc}$, for inelastic $pp$\/
collisions yielding one or more prompt long-lived charged particles 
in the kinematic range $\pt>0.2$\,\gevc\ and $2.0 <\eta<4.5$\/ is 
obtained using the expression

\vspace*{-0.5\baselineskip}
\begin{equation}
\label{eq:def}
     \sigma_{\rm inel}^{\rm acc}
    = \frac{I^{\rm acc}}{L} 
    = \frac{N^{\rm vis}}{\eps \cdot L} \;.
\end{equation}

\noindent
Here $I^{\rm acc}$\/ is the number of $pp$\/ interactions in data with a least
one prompt charged particle in the kinematic acceptance $\pt>0.2$\,\gevc\ and 
$2.0<\eta<4.5$\/ while  $L$\/ is the integrated luminosity of the data 
set under consideration. The number of interactions $I^{\rm acc}$\/ is proportional 
to the experimentally observed number of events, $N^{\rm vis}$, with at least
one reconstructed track in the fiducial region. The ratio $\eps=N^{\rm vis}/I^{\rm acc}$\/
is determined from the full simulation, which includes the possibility of multiple 
interactions per event, 

\vspace*{-0.5\baselineskip}
\begin{equation}
\label{eq:eps}
     \eps 
    = \frac{N^{\rm vis}_{\rm MC}}{I^{\rm acc}_{\rm MC}} 
    = \frac{N^{\rm vis}_{\rm MC}}{I^{\rm vis}_{\rm MC}}\cdot
      \frac{I^{\rm vis}_{\rm MC}}{I^{\rm acc}_{\rm MC}} \;.
\end{equation}

\noindent
The first factor, the ratio 
$N^{\rm vis}_{\rm MC}/I^{\rm vis}_{\rm MC}$\/ of events and interactions with at 
least one reconstructed track in the fiducial region, corrects for the 
fraction of multiple interactions. The second factor, the ratio 
$I^{\rm vis}_{\rm MC}/I^{\rm acc}_{\rm MC}$, is the efficiency to detect a single 
interaction with at least one prompt electron, muon, pion, kaon, proton 
or the corresponding antiparticle, in the kinematic acceptance.

To study the sensitivity of the analysis to the choice of the cuts on 
track quality and ${\rm DCA}$, the measurements are performed for two 
cases: ``loose'' settings accepting most reconstructed tracks, and 
``tight'' ones selecting mainly the cores of the $\chi^2/{\rm NDF}$\/ 
and ${\rm DCA}$ distributions.

\begin{figure}[t]
\centering
\includegraphics[width=0.475\textwidth]{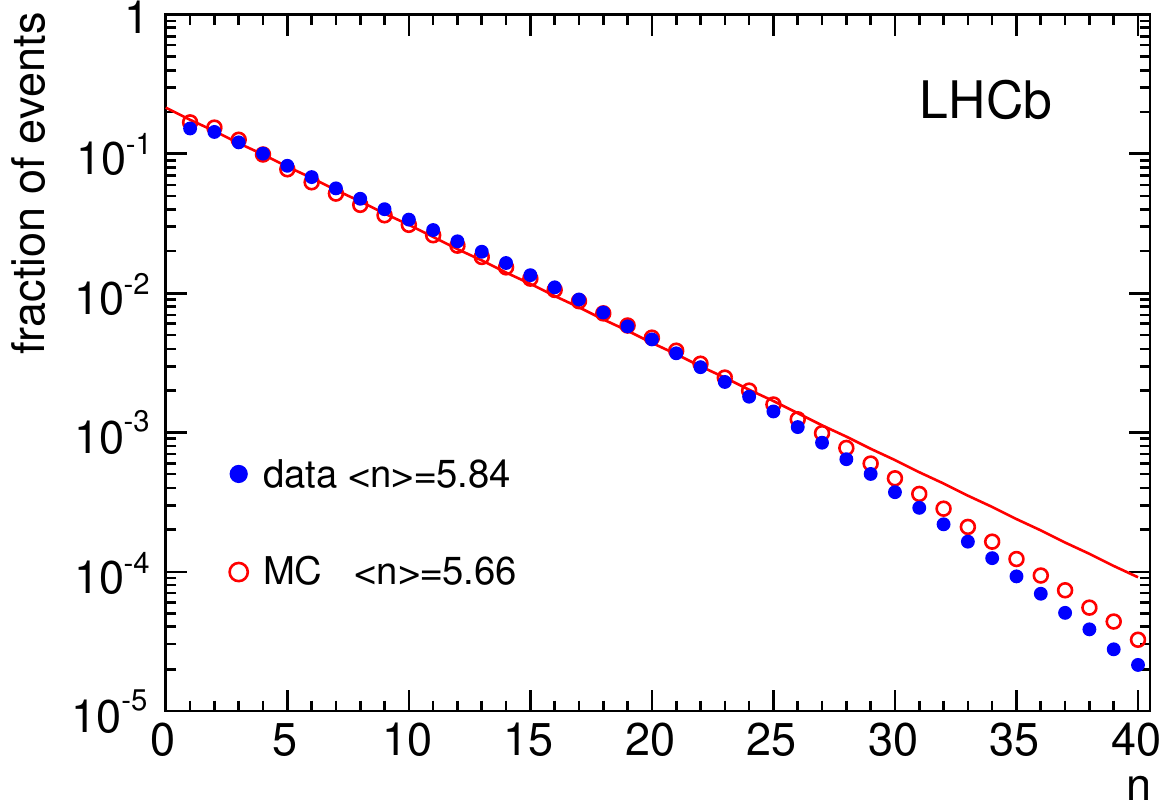}\hfill
\includegraphics[width=0.475\textwidth]{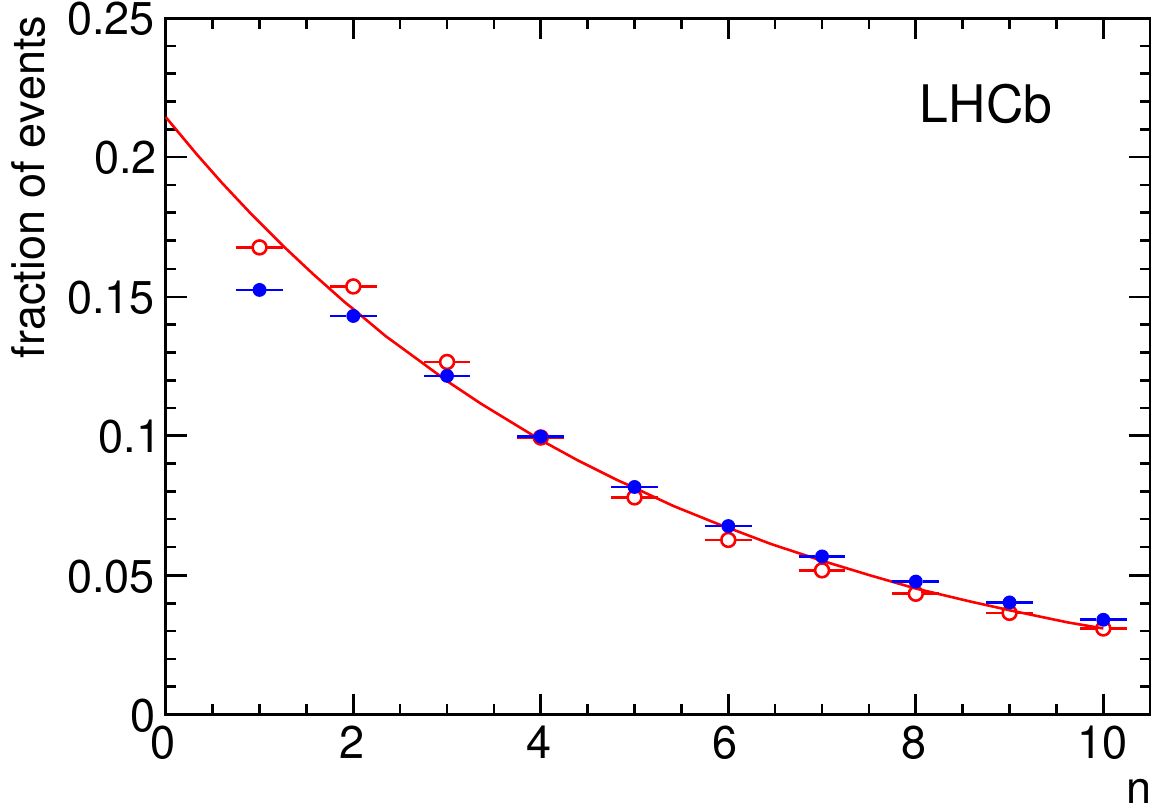}
\caption{\small
Normalized track multiplicity distributions with $n\geq 1$\/ tracks
in the fiducial region for the field-down configuration and tight cut 
settings in data and simulation. The superimposed function is an
exponential with the same average as the simulation. The right hand plot 
with a linear scale shows a zoom of the low-multiplicity region. The
vertical error bars are smaller than the symbol sizes. 
\label{fig:mult}}
\end{figure}

Figure\,\ref{fig:mult} shows the normalized multiplicity distributions of 
tracks from the luminous region that are recorded in the fiducial region
of the analysis for the tight cut settings in the 
field-down configuration. The distributions have an approximately 
exponential shape, as can be seen from the superimposed curves.
The small disagreement seen at low multiplicities is addressed when
discussing systematic uncertainties.

Table\,\ref{tab:cstab} gives the interaction and event counts in simulation 
and data. The simulations are based on a total of $I_{\rm MC}$\/ inelastic $pp$\/ 
interactions. The event counts in the simulation are given for an average 
of $P=1.05$\/ interactions per event and for both settings of the analysis cuts. 
One finds a typical value for the correction factor $\eps$ of 0.87.
For a given magnet polarity, the inelastic cross-section is taken to be
the central value of the measurements with loose and tight cuts. The final 
cross-section result is determined by the arithmetic average of the central 
values for the two magnet polarities. Here any biases that 
change sign under inversion of the field cancel exactly and uncertainties 
that are not fully correlated between the two configurations are reduced.
Within the acceptance of LHCb, the inelastic $pp$\/ cross-section with 
at least one prompt long-lived charged particle having $\pt>0.2\,\gevc$\/ and 
$2.0<\eta<4.5$\/ is found to be $\sigma_{\rm inel}^{\rm acc}=54.96\pm 0.01$\,mb,
where the uncertainty is purely statistical.

\begin{table}[t]
\centering
\caption{\small
Numbers of interactions and events, in multiples of $10^6$, in simulation and 
data for different magnetic field configurations and analysis cuts, and the 
resulting cross-sections in the kinematic acceptance. 
\label{tab:cstab}}
\begin{tabular}{lcc}
{\bf Simulation}                    & field-down & field-up \\       
\hline\\[-4mm]
$I_{\rm MC}$                           &  31.784  & ~\,4.948 \\
$I^{\rm acc}_{\rm MC}$                   &  26.121  & ~\,4.067 \\
$N^{\rm vis}_{\rm MC}$ (loose cuts)      &  22.907  & ~\,3.584 \\ 
$N^{\rm vis}_{\rm MC}$ (tight cuts)      &  22.693  & ~\,3.551 \\[2mm] 
{\bf Data}                   & &  \\
\hline\\[-4mm]
$N^{\rm vis}$(loose cuts)               &  30.098  &   60.285 \\ 
$N^{\rm vis}$(tight cuts)               &  29.735  &   59.541 \\[2mm] 

{\bf Cross-section [mb]}     & &  \\
\hline\\[-4mm]
 $\sigma^{\rm acc}_{\rm inel}$(loose cuts) &  55.36   & 54.73  \\    
 $\sigma^{\rm acc}_{\rm inel}$(tight cuts) &  55.20   & 54.55   
\end{tabular}
\end{table}

\section{Systematic uncertainties}
%----------------------------------
\label{systematic}
The systematic uncertainties are determined separately for the two 
magnet settings and are combined taking into account the correlations
between the individual contributions. The dominant uncertainty comes from 
the integrated luminosity, which is known with a precision of $3.5\%$. 
The sensitivity to the knowledge of the fraction of multiple interactions
was tested by varying $P$\/ in the simulation in the range 
$1.025\leq P \leq 1.075$, which leads to a variation in the 
cross-section of $1.5\%$. 

\begin{table}[t]
\centering
\caption{\small 
Summary of the relative systematic uncertainties, expresses as 
a percentage, for the 
measurement of the inelastic $pp$\/ cross-section measurement,  
separately for the two magnet polarities and the combined value.
\label{tab:syst}}
\begin{tabular}{lccc}               
Source & field-down  & field-up  & combined\\
\hline 
Luminosity                       &  3.5    &  3.5   &  3.5  \\ 
Multiple interactions            &  1.5    &  1.5   &  1.5  \\ 
Selection cuts                   &  0.3    &  0.3   &  0.3  \\
Calibration                      &  1.1    &  0.5   &  0.8  \\ 
Track finding efficiency         &  0.8    &  0.8   &  0.8  \\ 
Charged particle multiplicities  &  1.0    &  1.0   &  1.0  \\ 
Data taking period               &  1.0    &  1.0   &  0.7  \\ 
Azimuthal dependence             &  1.3    &  1.3   &  0.9  \\
Magnet polarity                  &  0.6    &  0.6   &  0.6     
\end{tabular}
\end{table}

Several systematic effects are related to a possible mismatch in the 
distributions of the selection variables between
data and simulation. The determination of the impact of the selection cuts 
on the event selection efficiency requires a proper modelling of the tails of 
the distributions of the selection variables. The corresponding systematic 
uncertainties are found to be $0.3\%$\/ by varying the selection cuts between 
loose and tight settings. The influence of the detector calibration on 
the reconstruction of charged tracks is tested by comparing the nominal event 
counts with those obtained when using an alternative version of the 
reconstruction code. For the loose cuts the changes are small, but for 
the tight cuts variations in the event counts of $1.1\%$\/ for field-down 
and $0.5\%$\/ for field-up are observed, which are assigned as systematic 
uncertainties. The systematic uncertainty on the reconstruction efficiency 
of a single track was found to be $3\%$ \cite{LHCb-PAPER-2010-002}. After 
convolution with the track multiplicity distribution of the events, this 
translates into an uncertainty of $0.8\%$\/ in the event selection efficiency.
The systematic uncertainty related to the modelling of the charged
particle multiplicity distribution in the kinematic acceptance is estimated from
the difference between the observed average multiplicities 
in data and simulation. At generator level the difference is about twice as
large, and a systematic uncertainty of $0.5$\/ units is assigned, which  
translates to a $1\%$ uncertainty in the event selection efficiency.

The cross-section measurement has been performed as a function of 
data taking period and in different azimuthal regions. Small but statistically 
significant variations are observed in both cases. From the maximum variations 
seen, uncertainties of $1.0\%$\/ and $1.3\%$\/ are assigned for dependencies 
on data taking period and azimuthal region, respectively. Finally, comparing 
the cross-section measurements for the field polarities one observes a 
difference of about $1.2\%$. Half of that variation is assigned as a 
systematic uncertainty.

The analysis has been performed in the LHCb laboratory frame which, 
due to a small crossing angle between the LHC beams, is slightly boosted
with respect to the $pp$\/ centre-of-mass system. It has been checked 
using simulation that this small boost has an impact of less than $0.1\%$\/ 
on the cross-section measurement. The contamination from elastic scattering 
events has been estimated to be negligible, and the statistical uncertainty 
due to the finite size of the Monte Carlo sample is less than $0.1\%$\/ 
and is neglected. Table\,\ref{tab:syst} gives a summary of the systematic 
uncertainties. For the combination of the two magnet polarities, the 
dependence on data taking period and the azimuthal dependence are 
assumed to be uncorrelated, while the other uncertainties are assumed to
be fully correlated. Adding the combined contributions in quadrature, 
the total systematic uncertainty on the cross-section is $4.3\%$.

\section{Results}
%----------------
\label{sec:Result}
The cross-section for inelastic $pp$\/ collisions at a centre-of-mass 
energy $\sqrt{s}=7$\,TeV, yielding one or more prompt long-lived charged 
particles in the kinematic range $\pt>0.2\,\gevc$\/ and $2.0<\eta<4.5$, is

\vspace*{-0.5\baselineskip}
\begin{equation*}
\sigma_{\rm inel}^{\rm acc}(\pt>0.2\,\gevc,\;2.0<\eta<4.5) = 55.0\,\pm\,2.4\;{\rm mb} \;,
 %%= 54.96\,\pm\,0.01\,({\rm stat})\,\pm\,2.30\,({\rm syst}) \;{\rm mb} \;,
\end{equation*} 

\noindent
with an uncertainty that is almost completely systematic in nature. 
The purely statistical uncertainty is two orders of magnitude smaller.

The measurement within the limited kinematic range above is scaled to 
full phase space with an extrapolation factor, $s_{\rm extr}$, which is given by the ratio 
of all inelastic interactions to the number of inelastic interactions within 
the kinematic acceptance. The \pythia 6 simulation used in the efficiency 
determination \cite{Sjostrand:2006za,LHCb-PROC-2010-056} gives
$s_{\rm extr}=I_{\rm MC}/I^{\rm acc}_{\rm MC}=1.2168\pm0.0001$, where 
the uncertainty is statistical. 

The extrapolation to full phase space is necessarily model dependent. 
To estimate its uncertainty, different soft QCD tunes provided 
by \pythia 8.201 (see Ref. \cite{Sjostrand:2014za} and references therein) 
have been considered: {\tt 4Cx}, a tune derived from the 
2C-tune to CDF data and adapted to LHC; {\tt Monash 2013}, a tune  
based on both $e^+e^-$ and LHC data; {\tt A2-CTEQ6L1}, {\tt A2-MSTW2008LO}, 
{\tt AU2-CTEQ6L1} and {\tt AU2-MSTW2008LO}, minimum bias and underlying
event tunes by the ATLAS collaboration using the CTEQ 6L1 and the 
MSTW2008 LO parton densities; and {\tt CUETP8S1-CTEQ6L1}, an underlying 
event tune by the CMS collaboration. Table \ref{tab:tunes} summarizes
some average properties of those tunes for non-diffractive, single-diffractive 
and double-diffractive interactions. Mean values and standard deviations
are given for $n$, the zero-suppressed average multiplicity of prompt long-lived
charged particles in the kinematic acceptance, for the visibility $v$, defined by 
the probability that at least one charged particle is inside the kinematic 
acceptance, and for the fraction $f$ of each interaction type. For any mix of 
interaction types, extrapolation factor and visibility are related by 
$s_{\rm extr}=1/v$.

The extrapolation factor, converting the inelastic cross-section in the 
kinematic acceptance to the total inelastic cross-section, is a function 
of the visibilities and the fractions of non-diffractive, single-diffractive 
and double-diffractive interactions. Since the interaction-type fractions are 
only weakly constrained by experiment (see e.g. Ref. \cite{ALICEcs:2013}), 
the values of $f$\/ given in Table \ref{tab:tunes} are not used in the
following. To determine an estimate for the uncertainty of the extrapolation
factor, a Monte Carlo approach is used. Multiplicities and visibilities are
generated according to Gaussian densities with parameters as given in 
Table \ref{tab:tunes}. The interaction type fractions that go into the 
extrapolation factor are then determined subject to the constraints that 
each of them lies between zero and one, that they sum to unity, and that 
the zero-suppressed average multiplicity of the mix is consistent with the
generator level average multiplicity of the \pythia 6 simulation, 10.93, which 
provides a good description of the data. The distribution of the average 
multiplicity is modelled according to a Gaussian function with this mean 
value and standard deviation $0.5$.

The method yields a distribution for $s_{\rm extr}$ with an average of $1.17$\/ and a 
standard deviation of $0.08$, which is assigned as the systematic uncertainty on 
the extrapolation factor obtained from the fully simulated Monte Carlo. The event 
fractions found by the above procedure, $0.70\pm0.12$, $0.17\pm 0.06$\/ and 
$0.13\pm 0.05$\/ for non-diffractive, single-diffractive and double-diffractive 
interactions, respectively, are consistent with the fractions given by the 
various tunes. The total inelastic cross-section becomes

\vspace*{-0.5\baselineskip}
\begin{equation*}
   \sigma_{\rm inel} = 66.9\,\pm\,2.9\,({\rm exp})\pm 4.4\,({\rm extr})\;{\rm mb} \;, 
\end{equation*} 

\noindent
with an experimental uncertainty (exp) that is dominated by systematic 
contributions and an extrapolation uncertainty (extr) of $7\%$.

\begin{table}[t]
\centering
\caption{\small Properties of soft QCD tunes in \pythia 8.201. For 
         non-diffractive, single-diffractive and double-diffractive 
         interactions, mean value and standard deviation over the 
         tunes considered in this study are given for average 
         multiplicities inside the kinematic acceptance, visibilities and 
         interaction type fractions.
\label{tab:tunes}} 
\begin{tabular}{cccc}
 interaction type   &  $n$   & $v$  & $f$ \\
\hline
 non-diffractive    &  $12.22  \pm 0.50$ & $0.9925 \pm 0.0003$ &  $0.713 \pm 0.002$ \\
 single-diffractive &  $~~5.94 \pm 0.29$ & $0.5059 \pm 0.0049$ &  $0.173 \pm 0.002$ \\
 double-diffractive &  $~~4.78 \pm 0.17$ & $0.5819 \pm 0.0062$ &  $0.114 \pm 0.001$
\end{tabular}
\end{table}

\begin{figure}[t]
\centering
\includegraphics[width=0.9\textwidth]{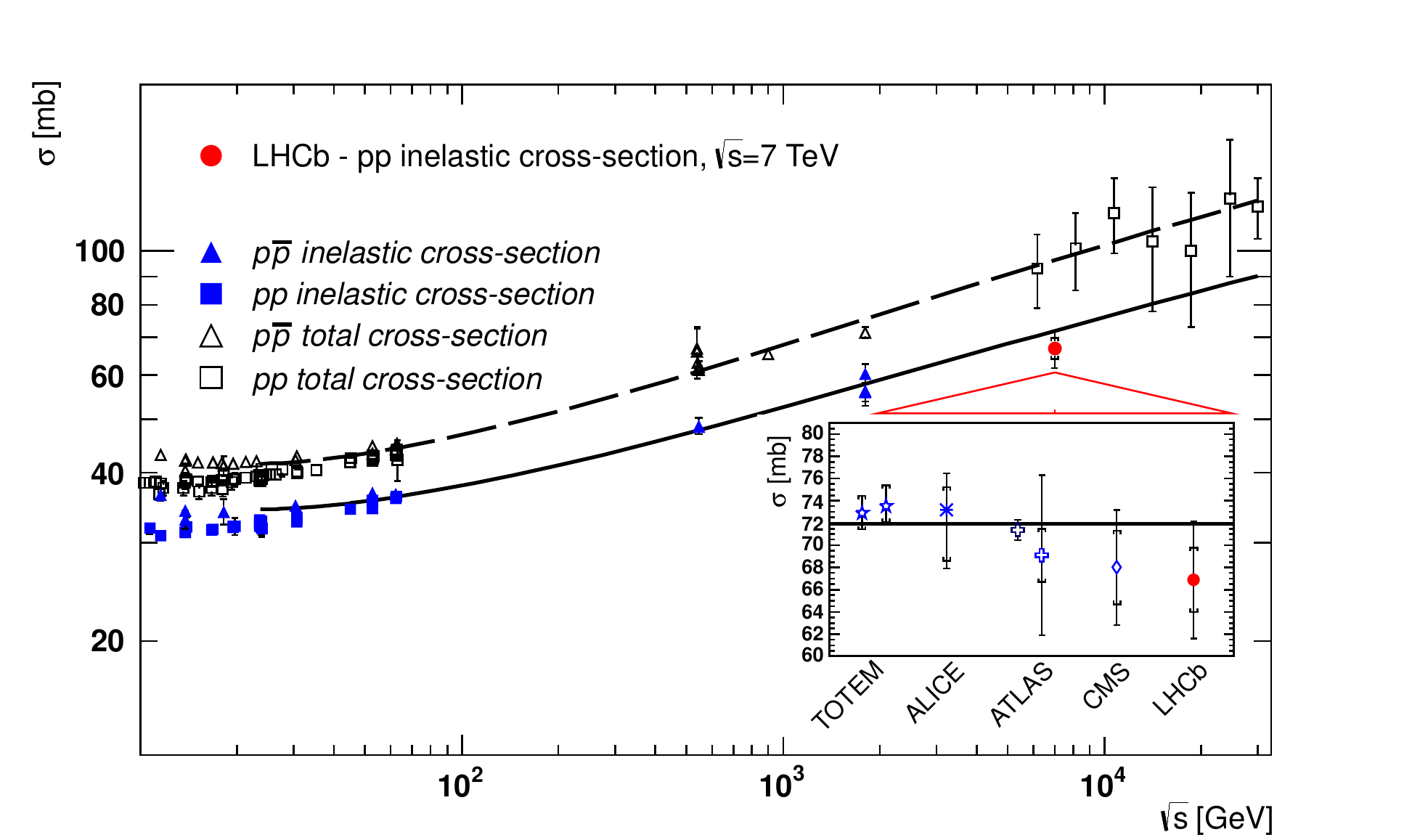}\hspace{1cm}
\caption{\small
Inelastic cross-section measured by LHCb compared to the existing 
data on the total {\protect{\cite{Nakamura:2010zzi}}} and inelastic 
cross-sections {\protect{\cite{Achilli:2011}}} in $pp$\/ and $p\bar{p}$\/ 
collisions as a function to the centre-of-mass energy. The full (dashed) 
line is a phenomenological fit {\protect{\cite{Fagundes:2013}}} of the 
energy dependence of the inelastic (total) cross-section. The main
plot only shows the LHCb measurement. The inset is a zoom, comparing 
all inelastic cross-section measurements by the LHC experiments
ALICE \cite{ALICEcs:2013}, ATLAS \cite{ATLAScs:2011,ATLASxs:2014}, 
CMS \cite{CMScs:2011} and TOTEM  \cite{TOTEMcs:2011,Antchev:2013iaa}.
The horizontal line represents the value of the phenomenological fit at
$\sqrt{s}=7$\,TeV. The error bars give the total uncertainties of the measurements.
When an inner error bar is shown, it represents the experimental uncertainties 
added in quadrature, while the full error bar also covers an extrapolation
uncertainty.
\label{fig:cspdg}}
\end{figure}

The LHCb result is displayed together with other cross-section measurements at 
various energies in Fig.\,\ref{fig:cspdg}. The data for the total 
cross-section are taken from Ref.\,\cite{Nakamura:2010zzi} and for the inelastic 
cross-section from Ref.\,\cite{Achilli:2011}. The plot shows that the available
measurements at centre-of-mass energies $\sqrt{s}>100$\,GeV can be described by 
a power-law behaviour. A $\ln^2s$\/ behaviour, as asymptotically expected if 
the Froissart-Martin bound is saturated, is not observed within the current 
experimental uncertainties. For comparison, results by the other LHC experiments
are also shown. The TOTEM \cite{TOTEMcs:2011,Antchev:2013iaa} and the 
ATLAS \cite{ATLASxs:2014} results are based on a measurement of the elastic 
cross-section, neither of which requires an extrapolation from a limited 
angular acceptance to full phase space. Within the extrapolation
uncertainties all results are in good agreement. Nevertheless, to avoid 
introducing ambiguities due to the model dependence of the extrapolation, any 
comparison between theory and the measurement presented in this paper
should be done for the restricted kinematic range $\pt>0.2\,\gevc$\/
and $2.0<\eta<4.5$.

\section*{Acknowledgements}

\noindent We express our gratitude to our colleagues in the CERN
accelerator departments for the excellent performance of the LHC. We
thank the technical and administrative staff at the LHCb
institutes. We acknowledge support from CERN and from the national
agencies: CAPES, CNPq, FAPERJ and FINEP (Brazil); NSFC (China);
CNRS/IN2P3 (France); BMBF, DFG, HGF and MPG (Germany);  INFN (Italy); 
FOM and NWO (The Netherlands); MNiSW and NCN (Poland); MEN/IFA (Romania); 
MinES and FANO (Russia); MinECo (Spain); SNSF and SER (Switzerland); 
NASU (Ukraine); STFC (United Kingdom); NSF (USA).
The Tier1 computing centres are supported by IN2P3 (France), KIT and BMBF 
(Germany), INFN (Italy), NWO and SURF (The Netherlands), PIC (Spain), GridPP 
(United Kingdom).
We are indebted to the communities behind the multiple open 
source software packages on which we depend. We are also thankful for the 
computing resources and the access to software R\&D tools provided by Yandex LLC (Russia).
Individual groups or members have received support from 
EPLANET, Marie Sk\l{}odowska-Curie Actions and ERC (European Union), 
Conseil g\'{e}n\'{e}ral de Haute-Savoie, Labex ENIGMASS and OCEVU, 
R\'{e}gion Auvergne (France), RFBR (Russia), XuntaGal and GENCAT (Spain), 
Royal Society and Royal Commission for the Exhibition of 1851 (United Kingdom).

\addcontentsline{toc}{section}{References}
\bibliographystyle{LHCb}
\bibliography{main,LHCb-DP,LHCb-CONF,LHCb-PAPER,LHCb-TDR}

\newpage
%%%%%%%%%%%%%%%%%%%%%%%%%%%%%%%%%%%%%%%%%%
\centerline{\large\bf LHCb collaboration}
\begin{flushleft}
\small
R.~Aaij$^{41}$, 
B.~Adeva$^{37}$, 
M.~Adinolfi$^{46}$, 
A.~Affolder$^{52}$, 
Z.~Ajaltouni$^{5}$, 
S.~Akar$^{6}$, 
J.~Albrecht$^{9}$, 
F.~Alessio$^{38}$, 
M.~Alexander$^{51}$, 
S.~Ali$^{41}$
% ...

{\footnotesize \it
% ...
$ ^{5}$Clermont Universit\'{e}, Universit\'{e} Blaise Pascal, CNRS/IN2P3, LPC, Clermont-Ferrand, France\\
$ ^{6}$CPPM, Aix-Marseille Universit\'{e}, CNRS/IN2P3, Marseille, France\\
% ...
$ ^{9}$Fakult\"{a}t Physik, Technische Universit\"{a}t Dortmund, Dortmund, Germany\\
% ...
$ ^{37}$Universidad de Santiago de Compostela, Santiago de Compostela, Spain\\
$ ^{38}$European Organization for Nuclear Research (CERN), Geneva, Switzerland\\
% ...
$ ^{41}$Nikhef National Institute for Subatomic Physics, Amsterdam, The Netherlands\\
% ...
$ ^{46}$H.H. Wills Physics Laboratory, University of Bristol, Bristol, United Kingdom\\
% ...
$ ^{51}$School of Physics and Astronomy, University of Glasgow, Glasgow, United Kingdom\\
$ ^{52}$Oliver Lodge Laboratory, University of Liverpool, Liverpool, United Kingdom\\
% ...
}
\end{flushleft}
%%%%%%%%%%%%%%%%%%%%%%%%%%%%%%%%%%%%%%%%%%

\end{document}